\documentclass[aps,pre,twocolumn,superscriptaddress,floatfix,nofootinbib,longbibliography]{revtex4-1}

\usepackage{yfonts}

\usepackage{mathrsfs}
\usepackage{amssymb}
\usepackage{amsmath}
\usepackage{amstext}
\usepackage[english]{babel}
\usepackage{graphicx}
\usepackage{bm}
\usepackage{float}
\usepackage{color}
\usepackage[breaklinks]{hyperref}
\hypersetup{colorlinks=true, linkcolor=blue, citecolor=blue, filecolor=blue, urlcolor=blue}

\makeatletter

\newcommand{\Rmnum}[1]{\expandafter\@slowromancap\romannumeral #1@}
\makeatother

\begin{document}

\title{Evaluating the Jones polynomial with tensor networks}

\author{Konstantinos Meichanetzidis}

\affiliation{School of Physics and Astronomy, University of Leeds, Leeds LS2 9JT, United Kingdom}

\author{Stefanos Kourtis}

\affiliation{Physics Department, Boston University, Boston, Massachusetts 02215, USA}

\date{\today}

\begin{abstract}
We introduce tensor network contraction algorithms for the evaluation of the Jones polynomial of arbitrary knots.
The value of the Jones polynomial of a knot is reduces to the partition function of a $q$-state anisotropic Potts model with complex interactions, which is defined on a planar signed graph that corresponds to the knot.
For any integer $q$, we cast this partition function into tensor network form, which inherits the interaction graph structure of the Potts model instance, and employ fast tensor network contraction protocols to obtain the exact tensor trace, and thus the value of the Jones polynomial. By sampling random knots via a grid-walk procedure and computing the full tensor trace exactly,
we demonstrate numerically that the Jones polynomial can be evaluated in time that scales subexponentially with the number of crossings in the typical case. This allows us to evaluate the Jones polynomial of knots that are too complex to be treated with other available methods. Our results establish tensor network methods as a practical tool for the study of knots.
\end{abstract}

\maketitle

\section{Introduction}

Knot theory is immensely interdisciplinary, with results and open questions spanning many fields of science, such as physics~\cite{KauffmanBook,witten1989,Qin2011,ENCISO201313,RICCA201321,MAGGIONI201329,LIU2013175,KnotsInPhysics}, quantum computation~\cite{Wocjan2006,pachos_2012,bremner,Goldberg2017}, quantum cryptography~\cite{QuantumMoney,PalumboCrypto}, chemistry and biology~\cite{Chemistry,DNAKnots,MolecularKnots,Mansfield1994}, study of every day life knotting of strands~\cite{Raymer16432}, and complexity theory~\cite{newalgounknot,NumOfReidemeister,KnottednessisinNP,ReidemeisterUpperBound}. A key notion in knot theory is that of a \textit{knot invariant} --- a quantity extracted from a knot $K$ which changes only under topology non-preserving knot operations, such as passing the knot through itself or cutting and recombining its strand. The \emph{Jones polynomial} $V_K(t)$~\cite{jones1985} --- a Laurent polynomial in $t\in\mathbb{C}$ --- is one such invariant that pervades knot theory. Knots $K, K^\prime$ are distinct if $V_K(t)\neq V_{K^\prime}(t)$. The Jones polynomial is thus pertinent to questions related to knottedness, such as the unknotting problem, a decision problem which is known to be in NP but unknown whether it lies in P~\cite{CompComplKnotLinkProb}. Hence, in addition to being central to the aforementioned applications of knot theory, evaluating the Jones polynomial is also a fundamental computational problem.

Exact evaluation of the Jones polynomial is generally a \#P-hard problem; computing $V_K(t)$ takes time that is expected to scale exponentially with the number of crossings in a knot. Exceptions to this occur for $t$ restricted to certain roots of unity, where $V_K(t)$ corresponds to quantum amplitudes of a quantum field theory~\cite{witten1989}, understood as braiding of anyons~\cite{KITAEV20032}. In particular, for $t=\pm1, \pm i, \pm e^{2\pi i/3}, \pm (e^{2\pi i/3})^2$, $V_K(t)$ can be evaluated efficiently~\cite{onthecompcomplofjonesandtutte}. Moreover, quantum algorithms can efficiently approximate the Jones polynomial at principal roots of unity in both the conventional quantum circuit model~\cite{Aharonov,Brennen} and the setting of topological quantum computation~\cite{RowellWang}. On the other hand, exponential classical algorithms that yield the \textit{full} expression for the Jones polynomial~\cite{EfficientAlgoColouredJones,KnotTheoryMathematica,SnapPy,KnotScape,Deguchi1,ELMISIERY1996249} in the general case have been implemented and are readily usable, but have a relatively small reach (up to $\sim20$ crossings).

Many knot invariants are intimately connected to statistical mechanical models~\cite{Wu92}. The Jones polynomial, in particular, is related to a $q$-state \textit{classical} Potts model with Anisotropic Complex Interactions (PACI).~\cite{thePottsmodel,bioPotts}.
Remarkably, the partition function $Z(q)$ of PACI, which is defined on an \textit{irregular planar graph} whose structure is defined by the topology of a knot $K$ is essentially the Jones polynomial $V_K(t(q))$ evaluated at
\begin{equation}
t(q) = \frac{1}{2} (q+\sqrt{q}\sqrt{q-4}-2) \,,\label{eq:potts2jones}
\end{equation}
up to a normalization.

Partition functions of classical models are of great interest in condensed matter physics and powerful algorithms have been developed to compute them, albeit mostly on graphs with periodic structure. Tensor network methods are an especially successful class of such techniques, which typically employ the renormalization-group procedure to efficiently approximate partition functions of classical lattice models very accurately~\cite{Levin2007a,Jiang2008,Gu2009,Xie2012,Evenbly2015,Zhao2016,Evenbly2017}. Recently, it was demonstrated that tensor network contraction schemes can also be very fast in obtaining the partition function of classical models \textit{exactly}, even on unstructured graphs with bounded degree and even when the underlying computation is a \#P-hard problem~\cite{fastcnting}.

In this work, we exploit the connection with statistical mechanics and the efficiency of tensor network methods to evaluate the Jones polynomial \textit{in the general \#P-hard case}. Specifically, we introduce tensor network contraction algorithms that can evaluate $V_K(t)$ at values of $t$ away from the ``easy'' ones, yet achieve demonstrably advantageous computation times that indicate subexponential scaling as a function of the number of crossings in $K$ for the \textit{typical} case. This affords us access to the value of the Jones polynomial of knots with 6 to 10 times as many crossings as what has been previously achieved in the literature with other methods. Our work thus furnishes a useful numerical tool for the evaluation of an essential knot invariant.

The rest of the paper is organised as follows.
In section~\ref{jones_potts} we review the mathematical connection
between the Jones polynomial and the partition function of the PACI model. We follow with section \ref{potts_tns} where
we present tensor network contraction methods for evaluating it
along with convincing numerical evidence of favourable resource scaling.
We conclude with section \ref{conclusion}.

\section{Jones polynomial evaluation as partition function of PACI}
\label{jones_potts}

We begin with a preliminary review of the relation between the PACI partition function and the Jones polynomial, starting with the relevant knot theory terminology. A knot $K$ consists of an embedding of the circle $S^1$ in $\mathbb{R}^3$. A knot diagram is the projection of the knot to $\mathbb{R}^2$, where the information about which strand is over which at every crossing $c$ is preserved. Intuitively, a knot diagram is what one produces when one attempts to draw a knot in two dimensions. Discarding the information about over and under crossing we obtain the knot \emph{shadow}.

\begin{figure}[t]
\centering
\includegraphics[width=\columnwidth]{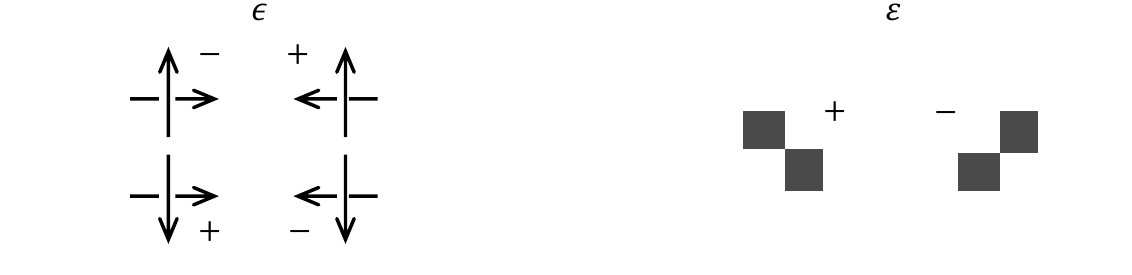}
\includegraphics[width=.73\columnwidth]{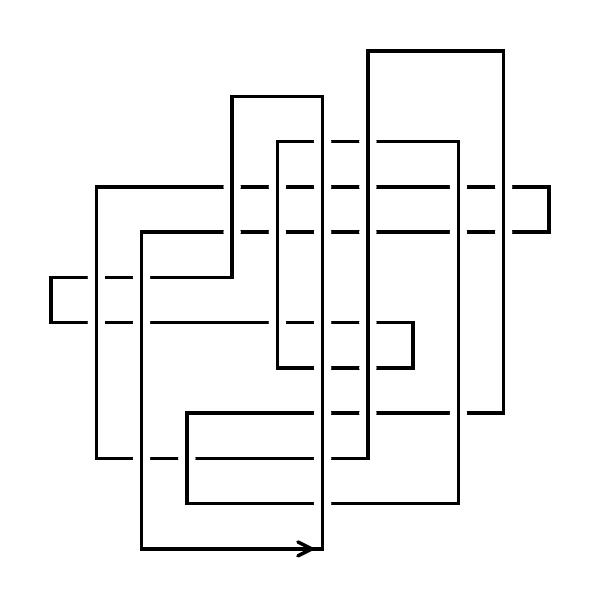}
\includegraphics[width=.73\columnwidth]{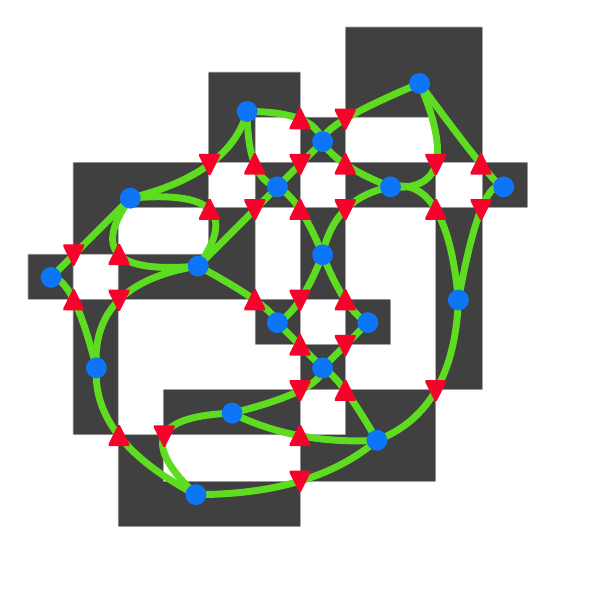}
\caption{ (Top) Twist (left) and Tait (right) signs for crossings of an oriented grid walk. (Middle) Oriented random knot diagram generated by a random grid walk for grid size $L=12$. (Bottom) Bicoloured knot diagram, its unsimplified Tait graph $G$ composed of vertices (blue dots) associated with black regions, and edges (green) decorated with $\varepsilon_c=+/-$ (up/down red triangles) associated with crossings.
The corresponding tensor network $G_\mathrm{T}$ comprises variable tensors (blue dots) connected through $q$-dimensional green lines with clause tensors (red triangles), pointing upwards (downwards) when representing a $J_+ (J_-)$ Potts interaction.
This knot is the right-handed Trefoil and for $q=3,5$ the Jones polynomial is $V(t(3))= i \sqrt{3}$, $V(t(5))= \frac{1}{2} (3 + \sqrt{5}) + (\frac{1}{2} (3 + \sqrt{5}))^3 - (\frac{1}{2} (3 + \sqrt{5}))^4$, as confirmed by our algorithm.}\label{fig:fig1}
\end{figure}

A knot can be oriented by choosing a direction along the strand. There are two ways to do this but they are equivalent. Each crossing obtains a \emph{twist sign} $\epsilon_c$ according to the direction of the strands exiting the crossing; if the strands cross in a clockwise (counterclockwise) fashion then the crossing obtains a positive (negative) twist sign (see Fig.\ref{fig:fig1}(top-left)). The sum of all the twist signs is called \emph{writhe}, ${w}_K=\sum_c \epsilon_c$, and characterises the knot chirality. The Jones polynomial is sensitive to the knot chirality as it can distinguish mirrored knots.

For any knot diagram, a planar graph $G$ called the \emph{Tait graph} is defined as follows. The two-dimensional regions defined by the knot diagram can be bicoloured with, say, black and white, so that no two adjacent regions share a colour. There are two ways to do this, and so we choose the convention that the unique unbounded region (background) is white.
In Fig.\ref{fig:fig1} we show an example of a knot K (middle) along with its bicolouring (bottom).
Then, vertices $v\in V$, where $V$ is the vertex set of $G$, correspond to the \emph{black regions}. Edges $c=(v,v^\prime)\in E$, where $(v,v^\prime) \in V\times V$ and $E$ is the edge set of $G$, are such that they connect black regions through the knot diagram crossings.
The graph $G$ thus obtained from $K$ is shown in Fig.\ref{fig:fig1} (bottom), where vertices are represented by blue dots and edges by green lines.
The vertex degree is denoted $d_v$ and counts the number of incident edges to that vertex. We have used the same symbol, $c$, for crossings and the corresponding edges. The edges are decorated by \emph{Tait signs} $\varepsilon_c$ which are determined by the following rule. If the region to the left (right) is black when exiting a crossing on the over strand, then the crossing obtains a positive (negative) Tait sign (see Fig.\ref{fig:fig1}(top-right)).
The sum of all tait signs is called \emph{Tait number}, $\tau_K=\sum_c \varepsilon_c.$
In Fig.\ref{fig:fig1} (bottom), the Tait signs are represented by red triangles decorating the edges, pointing up(down) for positive(negative).

We now restate the relation between the $q$-state PACI partition function $Z(q)$ and the Jones polynomial $V_K(t)$ of a knot $K$~\cite{zerosofjonesforfamiliesofknotsandlinks,zeroesofthejones}. A Potts model is placed on the Tait graph $G$ by defining spins with $q$ available states $\sigma_v=0,\dots,q-1$ to reside on the vertices $v=1,\dots,n_v$. The Tait signs $\varepsilon_c=\pm$ that decorate the edges of G determine the interaction strength between spins, which take two corresponding values $J_\pm\in\mathbb{C}$.
This rule which assigns interactions between the $q$-state spins renders the Potts model \textit{anisotropic}.
Their relation with the Jones variable is $e^{J_\pm}=-t^{\mp}$ and the Jones variable is determined by fixing $q\in \mathbb{N}$ via Eq.~\eqref{eq:potts2jones}. The Potts partition function over all spin-states $\{ \sigma \}$ is
\begin{align}
Z(q)  =&{\ } \sum_{\{\sigma\}} \prod_{(v,v^\prime)} T_{ \sigma_v \sigma_{v^\prime}} \,,\\
T_{ \sigma_v \sigma_{v^\prime}} =&{\ } 1-(1+t^{-\varepsilon_c})\delta_{\sigma_v  \sigma_{v^\prime}} . \label{eq:clause}
\end{align}
Multiplying the partition function $Z(q)$ with the appropriate prefactor $\mathcal{A}(q)$, which accounts for twists and ensures that the unknot returns $V_\bigcirc=1,~ \forall t\in\mathbb{C}$, we write
\begin{equation}
V_K(t(q)) = \mathcal{A}(q) Z(q)~, \label{eq:JonesPotts}
\end{equation}
$\mathcal{A}(q) = {(-t(q)^{\frac{1}{2}}-t(q)^{-\frac{1}{2}})}^{(-n_v-1)}  {(-t(q)^{\frac{3}{4}})}^{{w}_K} t(q)^{\frac{1}{4} \tau_K}$.

\section{Exact PACI partition function from tensor network contraction}
\label{potts_tns}

Our goal is the evaluation of $Z(q)$ of PACI at some $q\in\mathbb{N}$ of our choice in order to use Eq.\eqref{eq:JonesPotts}
and obtain the value of the Jones polynomial at $t(q)\in\mathbb{C}$ as defined in Eq.\eqref{eq:potts2jones}.
Note that computing the prefactor $\mathcal{A}(q)$ in Eq.\eqref{eq:JonesPotts} is in P as all $t(q)$, $w_K$, $\tau_K$ are efficiently computable. Thus one would focus on computing $Z(q)$ as efficiently as possible.

However, the evaluation of $Z(q)$ on arbitrary graphs is a \#P-hard problem~\cite{Geraci2008}. Regardless of this complexity, in this work we will use \textit{tensor network} methods~\cite{fastcnting} to obtain $Z(q)$ for Tait graphs $G$ \textit{exactly}. From a graph $G$ we construct a tensor network $G_\mathrm{T}$ encoding PACI as follows. Each vertex $v$ is endowed with a spin tensor (also known as a COPY tensor) of the form
\begin{equation}
\tilde{T}_{ \{ \sigma_v \}_{i=1}^{d_v} }= \delta_{ {\sigma_v}_1 {\sigma_v}_2 \dots {\sigma_v}_{d_v} } \,,
\end{equation}
which is a generalized $q^{d_v}$-dimensional Kronecker tensor. Each edge obtains a vertex on which we place the interaction tensor $T$ of Eq.~\eqref{eq:clause}, which is always a $q\times q$ matrix.
Reinterpreting Fig.\ref{fig:fig1} (bottom), we see the tensor network $G_T$ comprising $\tilde{T}$ and $T$ tensors
(blue dots and red triangles respectively).

\textit{Full contraction} of $G_\mathrm{T}$ yields $Z(q)$.
This amounts to performing a sequence of tensor contractions, each being a dot product over the common index of two adjacent tensors in $G_\mathrm{T}$
(green lines in Fig.\ref{fig:fig1} (bottom).
The partition function is then equivalently expressed as
\begin{equation}
Z(q)=\sum_{\{\sigma\}}\prod_{(\sigma_v,\sigma_{v^\prime})} T_{\sigma_v,\sigma_{v^\prime}} \prod_v \tilde{T}_{ \{ \sigma_v \}_{i=1}^{d_v} } \,.
\end{equation}
Every contraction step yields a \emph{graph minor} $H$ of the initial graph. Thus, when contracting a tensor network $G_\mathrm{T}$, there occurs at least one tensor of dimension equal to the maximal vertex degree over all minors $\Delta_H=\max_H \max_v d_v$.

\begin{figure}[t]
\centering
\includegraphics[width=\columnwidth]{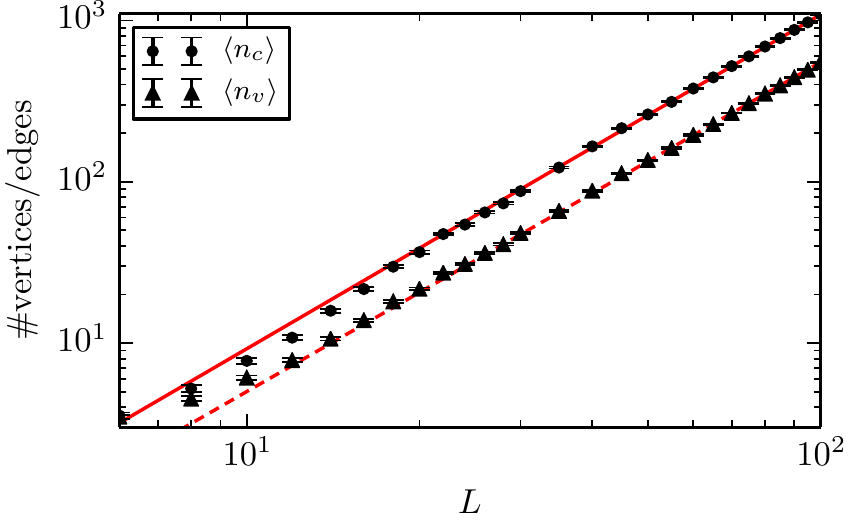}
\includegraphics[width=\columnwidth]{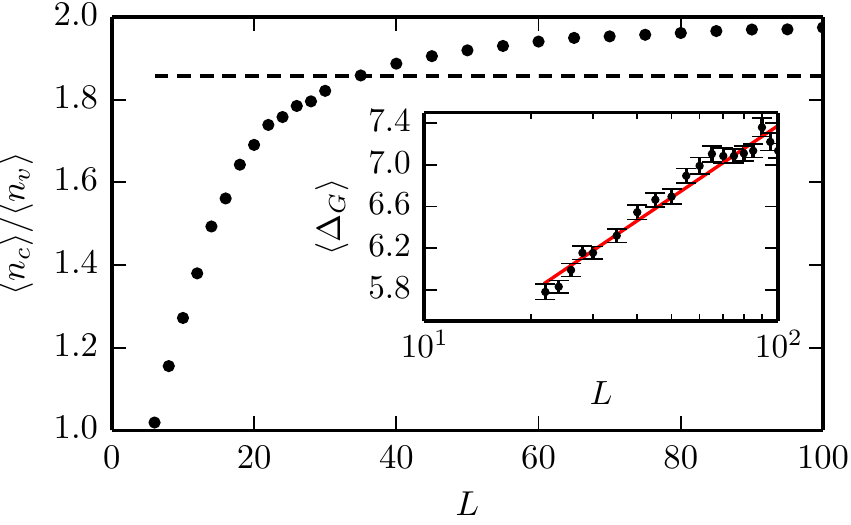}
\caption{(Top) Scaling of average number of vertices $\langle n_v \rangle$ (triangles) and edges $\langle n_c \rangle$ (dots) (logscale) of simplified Tait graphs with the the grid size $L$ (logscale) of the random grid walk. Dashed and solid red lines are linear fits on the last 15 data points with slopes $2.0132$ and $2.0742$, respectively. (Bottom) Ratio $\langle n_c \rangle$ over $\langle n_v \rangle$ versus $L$ converging to $\sim2$, compatible with the lower bound $\frac{13}{7}$ (dashed line) for random planar graphs. (Inset) Scaling versus $L$ (logscale) of average maximal degree $\langle \Delta_G \rangle$ of simplified Tait graphs $G$, for $L>22$. For each $L\in [6,100]$ we sampled 200 knots and error bars are standard mean error.} \label{fig:fig2}
\end{figure}

In general, finding the optimal contraction sequence so that $\Delta_H$ grows favourably slowly, is an NP-complete graph-theoretic problem~\cite{NPgraphs,NPgraphs1,NPgraphs2,Belmonte2013}. Practical contraction schemes for tensor networks is an active field of research~\cite{tncontraction}. We employ contraction methods introduced in Ref.~\cite{fastcnting}, where we refer the reader for explicit details of the methods,
here dubbed \texttt{greedy} and \texttt{METIS}, which were developed for fast evaluation of partition functions similar to $Z(q)$. In the \texttt{greedy} method, the ``cheapest'' edge contraction in terms of the resulting $\Delta_H$ is performed. On the other hand, \texttt{METIS} heuristically constructs a separator hierarchy using the METIS algorithm~\cite{Karypis1998}, attempting to minimize the cut length while splitting the graph into comparably large components. The contraction is performed following the separator hierarchy in a coarse graining fashion. For details we point to Ref.~\cite{fastcnting}. We provide an example script demonstrating the evaluation of $V_K(t)$ in an online repository~\cite{tensorcspgitlab}.

\subsection{Subexponential memory and runtime scaling}

To investigate the performance of our numerical scheme, we require a procedure for generating random knots, whose Tait graphs we can use to evaluate the Jones polynomial. Since Tait graphs are planar and connected by construction, one may be tempted to just generate random connected planar graphs. However, not any planar connected graph corresponds to a knot shadow. A generic planar connected graph corresponds to a link shadow, where a link is viewed as the embedding of multiple nonintersecting $S^1$ components. Instead, we employ the \emph{random grid walk} method~\cite{Cromwell1995,EvenZohar2017} to sample random knot diagrams, ensuring by construction that the number of components is always one. A grid walk consists of horizontal segments and vertical segments, where \emph{vertical segments always pass over horizontal ones}. The walk is encoded by a random permutation of coordinates $x,y\in S_L$, where $L$ is the linear grid size, and steps of the form $(x_i , y_i) \to (x_i,y_{i+1}) \to (x_{i+1},y_{i+1})$. Since all knots have a grid walk representation, any knot is accessible via this procedure. An example grid walk is shown in Fig.~\ref{fig:fig1} (middle).

For a given orientation of the grid diagram, each crossing has a twist sign. All possible configurations are shown in Fig.~\ref{fig:fig1} (top-left), and summing over them we obtain the writhe. The bicoloured knot along with its $G$ are shown Fig.~\ref{fig:fig1} (bottom). Keeping in mind that vertical segments pass over horizontal ones, the colour pattern around a crossing determines the Tait signs $\varepsilon_c$, as shown in Fig.~\ref{fig:fig1} (top-right), and so the Tait number is readily available.

With the random grid walk construction that allows us to randomly sample knots, we now investigate properties of the corresponding Tait graphs $G$. First, we perform Reidemeister moves that leave the knot topology invariant but simplify the graph. In particular, a Reidemeister I introduces or removes a twist in the knot. We employ it to remove loops, i.e. edges of the form $(v,v)$, as well as spikes, i.e,. degree-$1$ vertices, from the Tait graph. Note that a Reidemeister I move changes the writhe by $1$. A Reidemeister II move amounts to overlaying a strand over or under another, or inversely, combing the strands so they do not cross. These moves are usually referred to as poke and unpoke, respectively. In terms of $G$, we perform unpokes in order to remove double edges $c,c^\prime=(v,v^\prime)$ with $\varepsilon_c=-\varepsilon_{c^\prime}$. We then study the scaling of graph invariants of the resulting simplified graphs.

In Fig.~\ref{fig:fig2} (top) we provide evidence for quadratic scaling of average number of vertices, $\langle n_v \rangle\sim L^2$, and average number of edges, $\langle n_c \rangle\sim L^2$, for simplified Tait graphs $G$ obtained by the random grid walk. In Fig.~\ref{fig:fig2} (bottom) it is shown that the ratio of the average number of edges over the average number of vertices converges to $\sim2$ as the size of the Tait graphs increases. This convergent behavior is compatible with the lower bound $\frac{13}{7}$ of this ratio for random planar graphs~\cite{Gerke}. Furthermore, for random planar graphs the average maximal degree $\langle \Delta_G \rangle$, defined as $\Delta_G=\max_v d_v$, scales logarithmically with the graph size \cite{mcdiarmid_reed_2008}. This is also confirmed for the simplified Tait graphs sampled by the random grid walk, as shown in Fig.~\ref{fig:fig2} (inset).
Note that the data presented in Fig.~\ref{fig:fig2} are also a manifestation of the fact that the tensor networks, or equivalently, the interaction graphs of the PACI model relevant to the problem at hand, are irregular. This means that they are \textit{not} amenable to efficient methods that yield the Potts model's partition function on regular graphs, such as
the Corner Transfer Matrix Renormalization Group in the case of the square lattice.

\begin{figure}[t]
\centering
\includegraphics[width=\columnwidth]{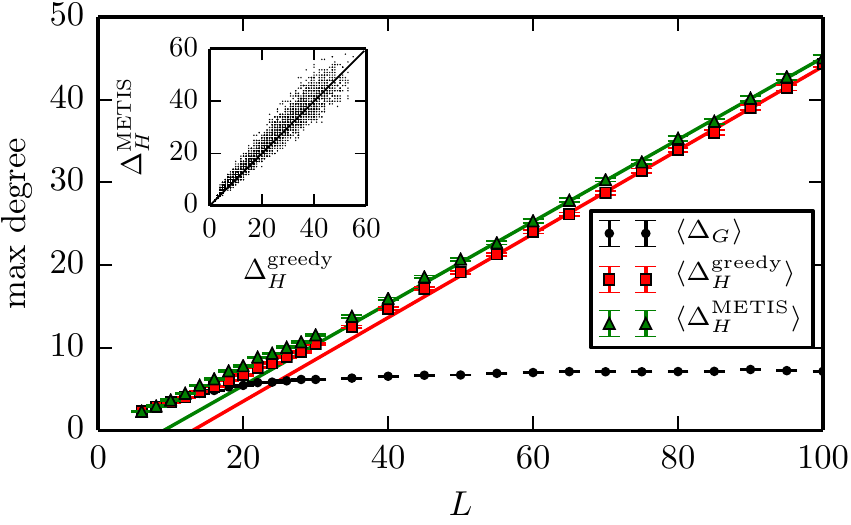}
\includegraphics[width=\columnwidth]{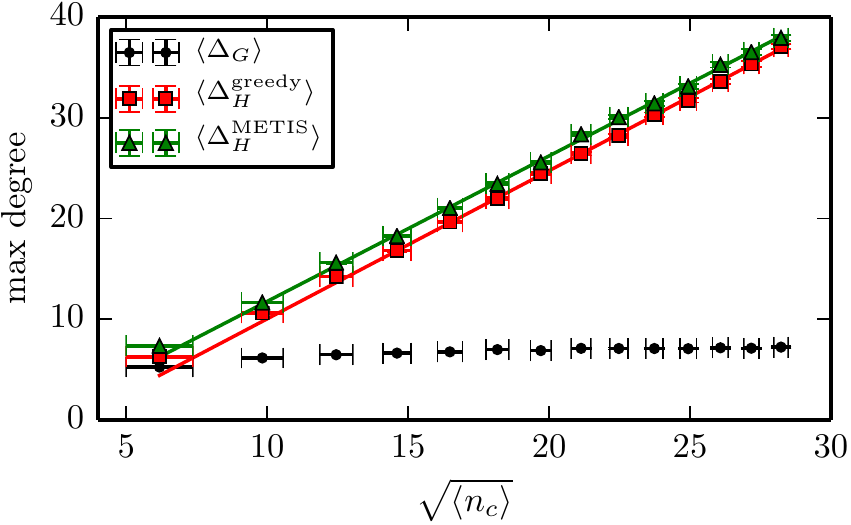}
\caption{Scaling with grid walk size $L$ (top) and with crossing number $\sqrt{n_c}$ (bottom) of average maximal degree $\langle \Delta_H \rangle$ encountered under \texttt{greedy} (red squares) and \texttt{METIS} (green triangles) contraction of $G$. Solid lines are linear fits (on the last 10 data points) showing the asymptotic behaviour. Black dots represent average maximal degrees $\langle \Delta_G \rangle$ of $G$. (Inset) Instance-by-instance comparison of $\Delta_H$ for the two contraction methods. The same data is used as for Fig.~\ref{fig:fig2}. Data points $\langle n_c \rangle$ are obtained by binning the interval between $\max n_c$ for $\min L$ and $\min n_c$ for $\max L$ and placing symbols at the mean of each bin. This is due to the fact that by sampling graphs for each $L$ we obtain a finite-variance distribution over $n_c$. Error bars are standard mean error.} \label{fig:fig3}
\end{figure}

\subsection{Numerical results}

The central quantity of interest for the purposes of tensor network contraction is the \emph{maximal degree} $\Delta_H$ encountered in the sequence of minors $H$ occurring during contracting $G$. This quantity characterizes the complexity of the algorithm, in the sense that runtime and memory requirements scale as $O(q^{\Delta_H})$.

\begin{figure}[t]
\centering
\includegraphics[width=\columnwidth]{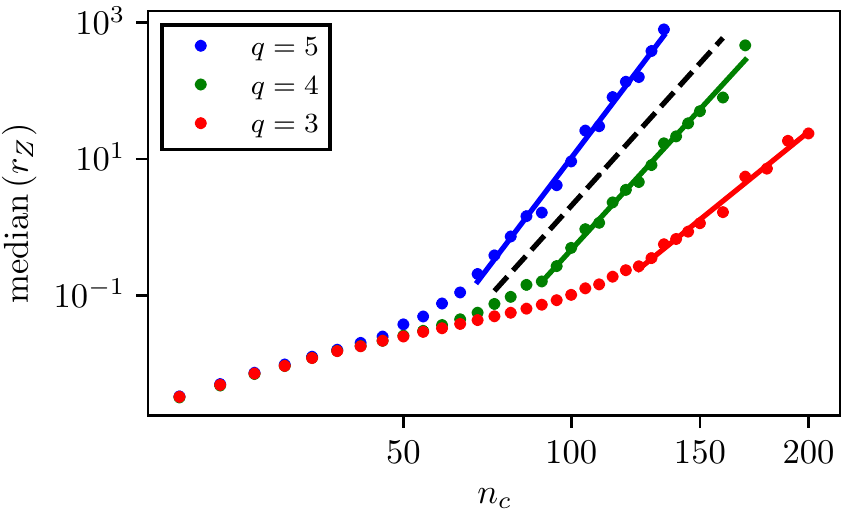}
\caption{Scaling of median runtime
$r_{Z(q)}$ for computing $Z(q)$ at $q=3,4,5$ via contracting $G_\mathrm{T}$ with \texttt{greedy}
as a function of $\sqrt{n_c}$
for typical instances:
For $n_c$ we sampled 1000 knots
by inverting the quadratic fit $\langle n_c \rangle(L)$ of Fig.\ref{fig:fig2} (Top) to obtain $L(\langle n_c\rangle)$ and sampling for the appropriate $L$ until 1000 knots were obtained.
The runtimes shown were computed for the graphs with $\Delta_H=\mathrm{median}(\Delta_H)$ for each $n_c$.
Dashed line of slope $0.93$ indicates the scaling of \texttt{TDTM}~\cite{Bedini_2010} and
the slopes for \texttt{greedy} are $0.68$, $0.92$, and $1.11$,
showing superiority for $q=3$ and matching performance for $q=4$.
Computations were performed on a single processor (Intel Xeon CPU E5-2667 0 \@ 2.90GHz) processor with $\sim$80~GB RAM.} \label{fig:fig4}
\end{figure}

In Fig.~\ref{fig:fig3} we show the scaling of the average $\Delta_H$ with the grid-walk size $L$. We find an asymptotically \emph{linear} scaling with $L$, which implies runtime scaling $O(q^{\sqrt{n_c}})$. Both contraction methods perform similarly, with \texttt{METIS} exhibiting marginally better scaling, yet only outperforming \texttt{greedy} for larger graphs ($n_c \gtrsim 900$).
We therefore use \texttt{greedy} to explicitly time the computation of $Z(q)$ for
realistically accessible graph sizes.
Runtime results for the cases of $q=3,4,5$ are shown in Fig.~\ref{fig:fig4}.

The favorable {\sl typical-case} scaling allows us to evaluate the Jones polynomial
for knots with $n_c=200$ for $q=3$
and with $n_c=135$ for $q=5$,
using moderate computational resources.
For comparison, the largest calculations of the {\sl full expression} for the Jones polynomial reported in the literature are for $n_c = 22$~\cite{EfficientAlgoColouredJones,KnotTheoryMathematica,SnapPy,KnotScape,Deguchi1}.
Furthermore,
for the {\sl exact} evaluation of $Z(q)$,
we compare our algorithm's $q$-dependent performance with that of the $q$-independent tree-decomposed transfer matrix algorithm (\texttt{TDTM})
 \cite{Bedini_2010},
which in the literature is presented for random planar
graphs of size up to $n_c=100$.
In Fig.~\ref{fig:fig4} we show that for $q=3$ the \texttt{greedy} tensor network algorithm outperforms \texttt{TDTM} for {\sl typical} instances.

The main bottleneck in these benchmarks is memory usage. For each $n_c$ there are exceptional knots that yield atypically large $\Delta_H$ and thus evaluation of $V_K(t)$ requires contraction of large tensors. With larger $q$, these exceptional cases may overflow the available memory, even though typical cases with the same $n_c$ are easily amenable.
On the other hand, for any particular knot of interest, one can test various graph contraction schemes to find the most favorable $\Delta_H$ and thus gauge the resources required a priori.
Then one can study typical cases alone, as we have demonstrated in Fig.~\ref{fig:fig4},
where for every $n_c$ we have obtained the runtimes for the graphs with $\Delta_H=\textrm{median}(\Delta_H)$.

Importantly, the asymptotic performance of our tensor network method does not depend on the content of the tensors, and so it is expected to perform as favorably for random planar instances of the Potts model, i.e. including those not corresponding to the Jones polynomial. Recall that Fig.~\ref{fig:fig2}
provides evidence that the graphs on which we have benchmarked
our tensor network algorithm can be considered as random planar graphs.
Therefore, and especially for the case of $q=3$,
this is a nontrivial result, as even incremental speedups in solving \#P-hard problems are rare.
We note that the slope change in the scaling of the median runtimes in Fig.~\ref{fig:fig4} is likely due to the absence of CPU cache misses when tensor sizes remain small throughout the contraction of the network. We therefore disregard small systems below this slope change when we obtain runtime scalings.

\section{Conclusions and outlook}
\label{conclusion}

In conclusion, we have developed a concrete methodology, based on tensor networks, for the evaluation of the Jones polynomial of arbitrary knots and demonstrated favorable performance of actual implementations.
Due to the broad relevance of knot invariants, our methods have wide applicability: classification of knotted polymers~\cite{Qin2011}, quantification of turbulence in classical and quantum fluids~\cite{RICCA201321}, and study of the Jones conjecture~\cite{JonesConj22cross} are just a few examples of problems that require computation of knot polynomials.
Furthermore, since Tait graphs are defined for links,
our algorithm trivially can be extended to the study of links, as well.
Recall that a link is the embedding of disjoint circles in $\mathbb{R}^3$ with the knot as a special case.
We therefore believe that the techniques introduced here can have multifaceted impact.

They also admit several extensions. For example, it is possible to obtain the coefficients of $V_K(t)$ via polynomial interpolation between evaluations at a number of values of $t$ equal to the degree of $V_K(t)$,
bounds to which are easily obtainable from the knot's bicolouring(which is efficient) and scale polynomialy with the number of crossings~\cite{KauffmanBook}.
Moreover, in analogy with condensed matter applications of tensor networks, where truncation of singular values along edges of the network lead to accurate approximations of a desired physical quantity, appropriate truncation procedures may allow one to obtain controlled approximations of the Jones polynomial, and potentially other knot invariants. It is also interesting to consider whether our algorithms can be extended to cases of $q\in \mathbb{R}$~\cite{Zqreal1,Zqreal}.
Indeed, recall that the evaluation of $V_K(e^{i 2\pi/n})$
corresponding to $q\leq4$ is a BQP-complete problem, except when $t=\pm1, \pm i, \pm e^{2\pi i/3}, \pm (e^{2\pi i/3})^2$
corresponding to $q=1,2,3,4$ for which the problem is in P.
This means that the quantum computation for the cases $q=1,2,3,4$ can be efficiently evaluated. Nevertheless, our algorithms are agnostic to the contents of the tensors and thus our results can extend to and be impactful for the study of the Potts model itself, as well as related graph theoretic problems such as $k$-colouring, where the interactions $J_{ij}$ decorating an interaction graph's edge $(i,j)$ need not be compatible with the topology of a link diagram.

\begin{acknowledgments}
We thank D.~Aasen, G.~Brennen, C.~Chamon, P.~Martin, A.~Michailidis, J.~Pachos, and Z.~Papic for comments on the manuscript and inspiring discussions.
S.K. was partially supported through the Boston University Center for Non-Equilibrium Systems and Computation. Preliminary numerical benchmarks were performed on the Boston University Shared Computing Cluster, which is administered by Boston University Research Computing Services.
K.M. acknowledges
the EPSRC Doctoral Prize Fellowship for financial support,
J. K. Pachos for providing a workstation at the School of Physics \& Astronomy, University of Leeds,
and B. Coecke for providing access to the \textswab{Duvel} server at the Department of Computer Science, University of Oxford, where runtime results where obtained.
\end{acknowledgments}

\clearpage

\end{document}